\documentclass{article}

\usepackage{comment}
\usepackage{enumitem}
\usepackage{booktabs}
\usepackage{physics}
\usepackage{float}
\usepackage{tikz}
\usepackage{tikz-3dplot}
\usetikzlibrary{arrows.meta,calc,positioning,3d}

\tikzset{
  plane/.style={fill=gray!6, draw=black!60, line width=0.5pt},
  axis/.style={->, line width=0.6pt},
  edgearrow/.style={-Latex, line width=0.9pt},
  dashedarrow/.style={-Latex, dashed, line width=0.8pt},
  centerpt/.style={circle, fill=black, inner sep=1.6pt},
  label/.style={font=\small}
}
\usepackage{graphicx}
\usepackage{amsmath}
\usepackage{amssymb}
\usepackage{cite}
\usepackage{url}
\usepackage[hidelinks]{hyperref}
\usepackage{geometry}
\providecommand{\keywords}[1]{%
  \smallskip\noindent\textbf{\textit{Keywords:}} #1
}

\title{
\textbf{MemoriesDB}: A Temporal-Semantic-Relational Database for Long-Term Agent Memory\\[0.5em]
\large\textit{Modeling Experience as a Graph of Temporal–Semantic Surfaces}
}

\author{
Joel ``val'' Ward\\
\textit{CircleClick Labs, Austin TX}\\
\texttt{val@ai.ccl.io}
}

\date{October 28, 2025}

\begin{document}

\makeatletter
\renewcommand\paragraph{
 \vspace{0.5em}
  \@startsection{paragraph}{4}{\z@}%
  {1ex \@plus 0.5ex \@minus .2ex}%
  {0.5ex \@plus .2ex}%
  {\normalfont\normalsize\bfseries}}
\makeatother

\maketitle
\vspace{-1.6em}

\begin{abstract}
We introduce \textbf{MemoriesDB}, a unified data architecture designed to avoid \emph{decoherence} across time, meaning, and relation in long-term computational memory.  Each memory is a \emph{time–semantic–relational entity}—a structure that simultaneously encodes \emph{when} an event occurred, \emph{what} it means, and \emph{how} it connects to other events.  Built initially atop PostgreSQL with \texttt{pgvector} extensions, \textbf{MemoriesDB} combines the properties of a \emph{time-series datastore}, a \emph{vector database}, and a \emph{graph system} within a single append-only schema.  Each memory is represented as a vertex uniquely labeled by its microsecond timestamp and accompanied by low-- and high--dimensional normalized embeddings that capture semantic context.  Directed edges between memories form labeled relations with per-edge metadata, enabling multiple contextual links between the same vertices.  Together these constructs form a \emph{time-indexed stack of temporal–semantic surfaces}, where edges project as directional arrows in a 1+1-dimensional similarity field, tracing the evolution of meaning through time while maintaining cross-temporal coherence.  This formulation supports efficient time-bounded retrieval, hybrid semantic search, and lightweight structural reasoning in a single query path. A working prototype demonstrates scalable recall and contextual reinforcement using standard relational infrastructure, and we discuss extensions toward a columnar backend, distributed clustering, and emergent topic modeling.

\end{abstract}

\keywords{vector database, time-series, graph database, agent memory, semantic retrieval, hybrid \mbox{architecture}, knowledge graph}

\section{Introduction}
Large language models (LLMs) have become the de facto substrate for modern artificial intelligence, yet they struggle to maintain long-term coherence. As their interactions extend over hours, days, or weeks, they suffer from what can be described as context decoherence: previously established facts and intentions drift out of scope, while the continuity of reasoning fragments into disjoint episodes. Existing architectures typically address this problem through sliding windows, retrieval-augmented generation (RAG), or episodic caches. While these techniques mitigate token limits, they do not provide a persistent substrate that encodes the temporal, semantic, and relational structure of an agent’s experience. Without such a substrate, continuity must be reconstructed ad hoc from text, and accumulated knowledge cannot be reasoned about as a coherent whole. They fail to preserve the evolving structure of experience that underlies reasoning and long-horizon planning.

\textbf{MemoriesDB}~\cite{ward2025} addresses this gap by treating memories as a first-class data system. It is a unified store in which every record—called a memory—is simultaneously a temporal event, a semantic vector, and a relational node (vertex). These three dimensions—time, meaning, and connection—form the core triality of the system. By integrating them into one append-only schema, MemoriesDB preserves both the sequence and the structure of experience, allowing agents to retrieve relevant information without losing the narrative thread that binds their actions together. Each labelled vertex represents a contextual embedding at a specific moment in time, while weighted edges capture relationship, strength, and confidence.

This design yields three key advantages.  It
\begin{enumerate}[nosep]
  \item enables \emph{persistent self-reference}: agents can recall and reason over their own past states without external indexing.
  \item provides a natural substrate for \emph{contextual inference}—by traversing similarity edges, an agent can reconstruct causal chains across time.
  \item scales pragmatically: updates occur asynchronously through \texttt{PubSub} channels, supporting distributed agents without retraining the core model.
\end{enumerate}

The result is an architecture that turns a stateless LLM into a continuous learning system capable of temporal reasoning and identity formation.
While not yet a complete cognitive model, this work represents a concrete step toward persistent, graph-grounded intelligence—an essential component in the path to artificial general intelligence (AGI).

\subsection{From fragments to coherence}

Human memory is not merely a sequence of tokens but a continuously evolving network of experiences. Coherence emerges from the interplay of three processes: (i) temporal ordering, which anchors experiences in chronology; (ii) semantic association, which links conceptually similar events; and (iii) structural relation, which encodes causal, conversational, or hierarchical connections. Most current memory systems capture only one of these axes. Vector databases represent meaning but not time or structure; time-series databases record sequence but not semantics; graph databases encode structure but lack a metric of similarity. MemoriesDB fuses all three into a single model where each stored experience can be queried across time, meaning, and relation simultaneously.

This design allows the system to represent memory as geometry. Each record occupies a position within a time-indexed graph of temporal–semantic surfaces. Within this geometry, the direction and weight of edges express how meaning propagates through time—preserving coherence across otherwise distant contexts. Queries can project along any axis or combination thereof: for example, “retrieve events semantically related to this idea within the past 24 hours,” or “find summaries that link these two topics.” The resulting framework treats recall as a navigation problem through a coherent spatiotemporal field rather than a static search in a single embedding space, thus escaping from the dreaded "curse of dimensionality" problem.

\subsection{Design principles}

The design of MemoriesDB follows four principles:

\begin{itemize}
    \item \textbf{Append-only architecture:} 
    All data are immutable once written. New memories extend the timeline rather than overwrite prior state. This approach ensures auditability, chronological integrity, and natural support for time-bounded queries.
    
    \item \textbf{Unified representation:} 
    Each memory is a typed vertex with normalized embeddings and JSON metadata. Edges are labeled, directional, and capable of carrying per-edge metadata, allowing multiple relations between the same vertices.
    
    \item \textbf{Compositional geometry:} 
    MemoriesDB models experience as a directed stack of 1+1-dimensional similarity fields—each vertex defines a local temporal–semantic plane. Edges project across these planes, forming a layered structure that preserves local continuity and global coherence.
    
    \item \textbf{Practical implementation:} 
    The current prototype runs on standard relational infrastructure (PostgreSQL with pgvector) and supports time-bounded recall, hybrid vector–SQL queries, and graph traversal. The system is designed for seamless migration to a columnar Parquet backend for distributed scaling.
\end{itemize}

\subsection{Contributions}

This paper makes the following contributions:
\begin{itemize}
    \item A unified data model that integrates time, semantic embeddings, and relational edges in a single append-only store.
    \item A geometric formulation of memory as a time-indexed graph of temporal--semantic surfaces, preserving coherence across long horizons.
    \item A working implementation demonstrating efficient time-bounded recall and hybrid semantic--structural queries on commodity SQL infrastructure.
    \item A research framework for analyzing long-horizon agent cognition, semantic drift, and emergent topic structure.
\end{itemize}

\subsection{Relation to prior work}

Prior research on vector databases, such as FAISS~\cite{faiss2019}, Milvus~\cite{milvus2024}, 
and pgvector~\cite{pgvector2024}, focuses on approximate nearest-neighbor retrieval in embedding space.
Time-series databases like InfluxDB and TimescaleDB~\cite{timescale2024} optimize for chronological queries
but lack semantic representations.
Graph databases, including Neo4j~\cite{neo4j2023} and TigerGraph~\cite{tigergraph2024},
specialize in topology but require external embedding layers to capture meaning.
\textbf{MemoriesDB} unifies these paradigms by embedding semantic vectors and graph relations directly
into a temporally ordered schema, enabling hybrid queries without cross-system joins.
Each addresses a single dimension of memory—semantic, temporal, or structural—but not all three simultaneously.

Beyond storage, \textbf{MemoriesDB} also relates conceptually to cognitive architectures such as 
ACT-R~\cite{anderson1998} and Soar~\cite{laird2012}, as well as retrieval-augmented generation (RAG)
approaches~\cite{lewis2020rag} and large models like GPT-4~\cite{openai2023gpt4}.
These systems highlight the importance of long-term structure, yet none provide a general-purpose
data substrate capable of maintaining high-dimensional embeddings, temporal metadata, and graph
relations under a single query model.

\subsection{Overview}

The remainder of this paper is organized as follows:
\begin{itemize}
    \item Section 2 describes the data model and geometric formulation of memory as time, meaning, and relation.
    \item Section 3 outlines the implementation on PostgreSQL and discusses query patterns for retrieval and reinforcement.
    \item Section 4 evaluates scalability and coherence retention under growing timelines.
    \item Section 5 explores future extensions—including a Parquet-backed columnar engine, distributed clustering, and automated topic discovery—and situates MemoriesDB within the broader pursuit of coherent, long-horizon cognition.
\end{itemize}

\section{Data Model and Geometry}
\label{sec:data-model}

MemoriesDB represents experience as a unified mathematical object that integrates temporal ordering, semantic representation, and relational connectivity.  
This section defines the core data structures, presents the geometric interpretation that underlies the system’s coherence, and outlines how retrieval operates within this space.

\subsection{The Memory Record}

Each record in MemoriesDB is a memory $M_i$ defined by the tuple
\begin{align}
M_i = (t_i,\, \kappa_i,\, \mathbf{V}_i,\, \mathbf{m}_i)
\end{align}
where
\begin{itemize}
    \item $t_i \in \mathbb{R}$ is the unique temporal coordinate of the memory;
    \item $\kappa_i$ is a categorical \emph{kind} label describing the record type (e.g., message, observation, summary, or state);
    \item $\mathbf{V}_i = \{\mathbf{v}_i^{(1)}, \mathbf{v}_i^{(2)}, \dots, \mathbf{v}_i^{(k)}\}$ is a collection of normalized sub-embeddings, each capturing a different representational view of the same memory (e.g., semantic, lexical, trigram, summary);
    \item $\mathbf{m}_i$ is a metadata map containing arbitrary key--value annotations (agent ID, topic tags, importance, etc.).
\end{itemize}

This formulation generalizes the notion of a single semantic vector to a
multi-view representation. Each sub-embedding $\mathbf{v}_i^{(n)}$
may differ in dimensionality, feature basis, or retrieval objective.\footnote{%
Here $n \in [0,k]$ indexes the available representational views
of each memory, with $\mathbf{v}_i^{(n)}$ denoting the $n$-th
sub-embedding (e.g., coarse, fine, or lexical).%
}

\makeatletter
\renewcommand\paragraph{
 \vspace{0.5em}
  \@startsection{paragraph}{4}{\z@}%
  {1ex \@plus 0.5ex \@minus .2ex}%
  {0.5ex \@plus .2ex}%
  {\normalfont\normalsize\bfseries}}
\makeatother

\paragraph{Practical instantiation}

In the current prototype, $\mathbf{V}_i$ typically contains both low- and
high--dimensional semantic embeddings $(\mathbf{v}_i^{(L)}, \mathbf{v}_i^{(H)})$,
with optional lexical or trigram projections.\footnote{%
The notation $(\mathbf{v}_i^{(L)}, \mathbf{v}_i^{(H)})$ is used informally to
denote coarse and fine representational views. Additional pseudo-embeddings such
as BM25~\cite{robertson2009bm25}, trigram overlap~\cite{cavnar1994ngrams}, or
other domain-specific projections can be included in $\mathbf{V}_i$ as needed.%
}

\paragraph{Retrieval function}
From a retrieval perspective, the system treats each memory as
\begin{align}
M_i = (t_i,\, \kappa_i,\, \mathbf{v}_{\mathrm{fuse},i},\, \mathbf{m}_i)  
\end{align}
where the fusion operator $f_{\mathrm{fuse}}$ aggregates the sub-embeddings
\begin{align*}
\mathbf{v}_{\mathrm{fuse},i} = f_{\mathrm{fuse}}(\mathbf{V}_i)
\end{align*}
The function $f_{\mathrm{fuse}}$ may implement a weighted combination,
Reciprocal Rank Fusion (RRF)~\cite{cormack2009rrf},
or another hybrid ranking strategy. This approach provides flexibility to
integrate new embedding types without altering the storage schema.

\vspace{0.5em}
This multi-view memory representation forms the foundation for the distance
and coherence metrics introduced in Section~\ref{sec:distance}.

\subsection{Edges and Relations}

Directed relations between memories are expressed as labeled edges:
\begin{align}
E_{ij} &= (M_i \rightarrow M_j,\; \rho_{ij}\,,\; W_{ij},\, \mathbf{m}_{ij})
\end{align}
where $\rho$ is a relation label (e.g., \textit{reply}, \textit{summary-of}, \textit{related-to}),  
\(W_{ij}\) is weight,  
and \(\mathbf{m}_{ij}\) is a metadata map attached to that specific edge.  
Multiple labeled edges may exist between the same vertex pair, forming a directed multigraph. Furthermore, W is defined as
\[
W = (w_{\mathrm{strength}},\, w_{\mathrm{confidence}})
\]

Each edge originates from the source plane of \(M_i\) and projects to the origin of its destination \(M_j\).  
This projection defines a local vector
\begin{align*}
\mathbf{e}_{ij} = (\Delta t_{ij},\, s_{ij})
\end{align*}
with \(\Delta t_{ij} = t_j - t_i\) (temporal displacement) and  
\(s_{ij} = 1 - \cos(\mathbf{v}_i^{(H)},\, \mathbf{v}_j^{(H)})\) (semantic displacement).  
The collection of all outgoing edges from a memory constitutes its local flow field \(\mathcal{F}_i\).

\subsection{The Temporal--Semantic Stack}

Because no two memories share precisely identical timestamps, time serves as a discrete indexing axis.  
Each memory therefore owns a unique local coordinate plane \(\mathcal{P}_{t_i}\) parameterized by \((\Delta t,\, \Delta s)\).  
The global structure of the database is the ordered stack
\begin{align*}
\mathcal{P} = \bigcup_{i=1}^{N} \mathcal{P}_{t_i}
\end{align*}
where edges form directed connections between planes.  
Intuitively, the system resembles a laminated sheet of temporal--semantic 
surfaces linked by arrows that trace relationships of meaning across time; 
collectively, the structure reveals how semantic organization evolves (see Figure 1).

This stack preserves cross-temporal coherence: semantic relationships are embedded directly in the temporal ordering rather than reconstructed from text.  
When an agent revisits a topic after a long interval, the retrieval path naturally bridges earlier layers of the stack, re-establishing continuity.

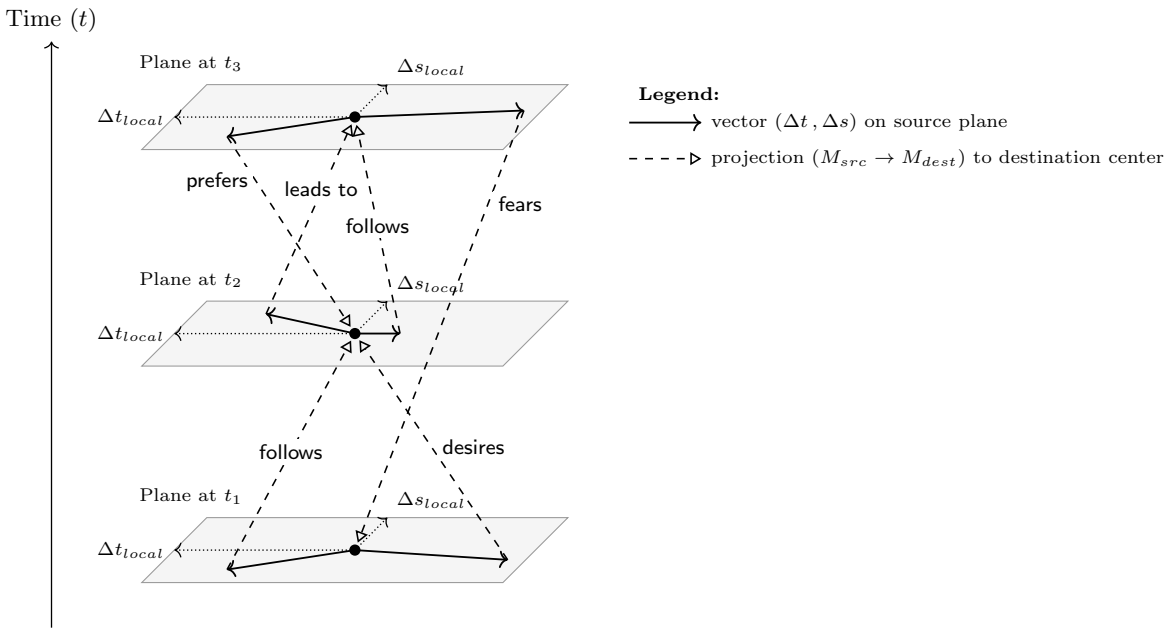
\begin{figure}[ht]
\centering
\begin{tikzpicture}[scale=1.2]
    \tikzset{
        plane/.style={fill=gray!8, draw=black!40, line width=0.4pt},
        axis/.style={->, line width=0.5pt},
        tick/.style={line width=0.4pt},
        centerpt/.style={circle, fill=black, inner sep=1.5pt},
        edgearrow/.style={->, line width=0.7pt},
        dashedarrow/.style={-{Triangle[open]}, dashed, line width=0.6pt},
        label/.style={font=\scriptsize}
    }

    \newcommand{\memplane}[5]{%
        \coordinate (P#5A) at (#1,#2);
        \coordinate (P#5B) at ($(#1+#3,#2)$);
        \coordinate (P#5C) at ($(#1+#3+#4*0.6,#2+#4*0.6)$); 
        \coordinate (P#5D) at ($(#1+#4*0.6,#2+#4*0.6)$);
        \filldraw[plane] (P#5A) -- (P#5B) -- (P#5C) -- (P#5D) -- cycle;
        \path let \p1 = ($(P#5A)!0.5!(P#5C)$) in coordinate (C#5) at (\p1);
        \node[centerpt] at (C#5) {};
        \draw[axis, densely dotted] (C#5) -- ($(P#5C)!0.5!(P#5D)$) coordinate (midAB) node [label,above right] {$\Delta s_{local}$};
        \draw[axis, densely dotted] (C#5) -- ($(P#5A)!0.5!(P#5D)$) coordinate (midAD) node [label,left ] {$\Delta t_{local}$};

        \node[label, anchor=south east] at ($(P#5D)+(0.5,0.05)$) {Plane at $t_{#5}$};
    }

    \def\W{4.0}  
    \def\D{1.2}  
    \def\DY{2.4} 

    \draw[axis] (-0.4,-0.5) -- (-0.4,\DY*2.5) node[above] {\small Time ($t$)};

    \memplane{0.6}{0}{\W}{\D}{1}
    \memplane{0.6}{\DY}{\W}{\D}{2}
    \memplane{0.6}{\DY*2}{\W}{\D}{3}

    \coordinate (S1) at (C1); 

    \coordinate (T2) at (C2);
    \coordinate (T3) at (C3);

    \newcommand{\vecOnPlane}[6]{%
        \pgfmathsetmacro\u{0.5 + #1*0.5}
        \pgfmathsetmacro\v{0.5 + #2*0.5}

        \coordinate (TMP#3#5) at
            ($ (P#3A)!\u!(P#3B) + (P#3A)!\v!(P#3D) - (P#3A) $);

        \draw[edgearrow] (C#3) -- (TMP#3#5);
        \draw[dashedarrow, shorten >=3pt] (TMP#3#5) -- (C#4)
            node[midway, fill=white, font=\footnotesize\sf, inner sep=2pt, #6]{#5};
    }

    \vecOnPlane{0.25}{0.0}{2}{3}{follows}{centered}
    \vecOnPlane{-0.6}{0.6}{2}{3}{leads to}{centered, xshift=4pt, yshift=10pt}
    \vecOnPlane{-0.6}{-0.6}{1}{2}{follows}{centered}
    \vecOnPlane{0.9}{-0.3}{1}{2}{desires}{right, xshift=2pt}
    \vecOnPlane{0.9}{0.2}{3}{1}{fears}{right, yshift=48pt, xshift=20pt}
    \vecOnPlane{-0.6}{-0.6}{3}{2}{prefers}{left, yshift=20pt, xshift=-14pt}

    \node[label, anchor=west] at (\W+2.0, \DY*2+0.6) {\textbf{Legend:}};
    \draw[edgearrow] (\W+2.0, \DY*2+0.3) -- ++(0.8,0) node[label, right] {vector ($\Delta t\,, \Delta s)$ on source plane};
    \draw[dashedarrow] (\W+2.0, \DY*2-0.1) -- ++(0.8,0) node[label, right] {projection ($M_{src} \to M_{dest}$) to destination center};

\end{tikzpicture}

\caption{%
    \textbf{Time-indexed stack of temporal--semantic planes.}
    Each memory at time~$t_i$ defines a local plane parameterized by $(\Delta t,\,\Delta s)$.
    Directed edges (solid) encode temporal--semantic offsets on the source plane,
    while cross-plane projections (dashed) trace their relationships to memories at earlier or later times.
    Together, the planes visualize how meaning and influence propagate coherently through time.
}

\label{fig:temporal_semantic_stack_tikz}
\end{figure}

\subsection{Distance and Coherence Metrics}
\label{sec:distance}

We begin with an idealized view of temporal–semantic geometry. For two memories $M_i$ and $M_j$, let us define the
local displacement vector:
\begin{align*}
\boldsymbol{\delta}_{ij} = \langle \lambda_t\,\Delta t_{ij}\,,\; \lambda_s\,s_{ij} \rangle
\end{align*}
where $\Delta t_{ij}$ is the temporal separation between memories
$M_i$ and $M_j$, and $s_{ij}$ is their semantic displacement.
The coefficients $\lambda_t$ and $\lambda_s$ balance the relative influence
of the temporal and semantic change.
Intuitively, $\boldsymbol{\delta}_{ij}$ points through the memory graph
from one vertex to another, tracing the local trajectory of meaning through time and experience.

The magnitude of the idealized distance function captures how far the memory has moved in the combined
temporal–semantic field:
\begin{align*}
d(M_i\,, M_j) = \sqrt{(\lambda_t \, \Delta t_{ij})^2 + (\lambda_s \, s_{ij})^2}
\end{align*}
where \(\lambda_t\) and \(\lambda_s\) weight time and meaning, respectively.  
This metric defines the local curvature of the temporal--semantic field.  
Low curvature indicates stability of topic or intent; high curvature signals conceptual drift.

\paragraph{Practical form}
Computing Euclidean norms require both square and square root operations, which introduce nontrivial computational overhead, especially in high--dimensional or real-time settings. Therefore, in practice, the magnitude is computed as follows:
Given two memories $M_i$ and $M_j$, their semantic distance is defined over the
fused representations of their respective embeddings:
\begin{align}
d(M_i\,, M_j) &= \left\| f_{\mathrm{fuse}}(\mathbf{v}_i^{(H)}) - f_{\mathrm{fuse}}(\mathbf{v}_j^{(H)}) \right\|_2
\label{eq:distance}
\end{align}

This expression measures similarity in the unified semantic space produced by
$f_{\mathrm{fuse}}$, which may internally combine multiple representational
channels (e.g. coarse and fine semantic vectors, lexical features, or trigram
signals). The fusion operator $f_{\mathrm{fuse}}$ may correspond to a learned weighting
scheme, a normalized linear projection, or a rank-fusion operator such as
Reciprocal Rank Fusion (RRF)~\cite{cormack2009rrf}.

During retrieval, $f_{\mathrm{fuse}}(\mathbf{V})$ blends coarse ($\mathbf{v}^{(L)}$)
and fine ($\mathbf{v}^{(H)}$) embeddings and/or additional lexical signals. However, for computing
pairwise distances in coherence and drift analysis, only the high--dimensional component
$\mathbf{v}^{(H)}$ is used, providing a more stable and semantically precise measure.

\vspace{1em}
\textit{Implementation note:}
MemoriesDB stores fused vectors in \texttt{pgvector} and performs
retrieval using the inner-product operator (\texttt{<\#>}), which avoids
square-root and normalization costs associated with true Euclidean distance.
When the vectors are unit-normalized at insertion time, this operator is
equivalent to cosine similarity, providing a fast and monotonic proxy for
semantic distance.

\vspace{1em}

\paragraph{Coherence metrics}

Pairwise coherence between two memories is measured as
\begin{align*}
C_{\mathrm{pair}}(M_i\,, M_j) = e^{-d(M_i, M_j)}
\end{align*}
yielding a scalar in $(0,1]$ that decays with both temporal and semantic separation.
The distance function $d(\cdot)$ is computed within the fused high--dimensional
embedding space $\mathbf{v}^{(H)}$.

Aggregating pairwise coherence over edges provides a local measure of the system’s temporal--semantic continuity.
To evaluate the structural stability of memories over duration, MemoriesDB maintains a
time-varying coherence signal. For an active window of edges
$E_t = \{(i,j)\}$ within a temporal interval $[t-\Delta t,\, t]$, \emph{local coherence} is
defined as
\begin{align}
C_{\mathrm{local},\,t} &= \frac{1}{|E_t|} \sum_{(i,j)\in E_t} e^{-d(M_i,M_j)}
\label{eq:coherence}
\end{align}

Higher $C_{\mathrm{local},\,t}$ values indicate strong temporal--semantic consistency across
adjacent memories, while lower values signify drift or contextual divergence.
This coherence measure also functions as a proxy for \emph{relative importance}:
memories that maintain high pairwise similarity over time are reinforced during
retrieval, whereas low-coherence regions become candidates for summarization or
decay.

Empirically, maintaining a high coherence \(\mathcal{C}\) correlates with improved recall relevance in long-horizon agents.

\subsection{Graph Geometry}

The overall data structure can be regarded as a directed multigraph \(G = (V,E)\) embedded in a product space
\begin{align*}
\mathbb{R}_t \times \mathbb{R}^{d_H}_v,
\end{align*}
augmented with discrete relational labels.  
This embedding produces a fibered graph: time acts as the base coordinate, while each vertex's local semantic--relational fiber contains its outgoing edges.  
Traversing the graph along increasing \(t\) yields a path of semantic transformations---analogous to the trajectory of a thought through conceptual space.

Edges projected onto their source planes define a local vector field  
\(\mathcal{F} = \{\mathcal{F}_i\}\).  
Integrating these local fields reconstructs the agent’s global semantic trajectory, providing a geometric interpretation of coherence over time.

\subsection{Query Semantics}

A query in MemoriesDB specifies constraints along one or more axes:
\begin{itemize}
    \item \textbf{Temporal window:} \([t_{\text{min}},\, t_{\text{max}}]\);
    \item \textbf{Semantic vector:} \(\mathbf{q}\) (embedding of query text);
    \item \textbf{Relational filter:} labels or metadata conditions.
\end{itemize}
The engine evaluates the query by:
\begin{enumerate}
    \item Restricting to records within the time window;
    \item Ranking candidates by similarity \(\text{sim}(\mathbf{v}_i^{(H)},\, \mathbf{q})\);
    \item Optionally expanding through outgoing edges within a coherence radius \(C \geq \tau\);
    \item Re-ranking by a combined importance score
    \[
    S_i = \alpha\,\text{sim}(\mathbf{v}_i^{(H)},\mathbf{q}) + 
          \beta\, e^{-\Delta t_i / \tau} + 
          \gamma\,\Phi_i,
    \]
    where \(\Phi_i\) encodes local edge density or relation type.
\end{enumerate}
This process unifies time-bounded search, semantic similarity, and structural reasoning in a single pipeline.

\subsection{Storage Realization}

Although the formalism is model-agnostic, the current prototype implements the data model using relational tables with vector and JSON fields.  
The append-only design ensures immutability and supports efficient partitioning by user and time.  
Indexes on both time and vector fields enable near-linear scan performance for moderate-scale agents.  
Edges are stored as independent rows with \texttt{(source, destination, relation, weight, meta)} columns, providing full multigraph support and hash-addressable metadata.

\subsection{Interpretation}

From a systems perspective, MemoriesDB functions as a \emph{coherence engine}.  
By embedding semantic relationships directly into the temporal order, it prevents the context fragmentation that typically accompanies long-horizon reasoning.  
From a cognitive perspective, the database approximates an episodic memory that maintains phase alignment between experience and meaning---analogous to preventing information-theoretic decoherence in quantum systems.  
The resulting architecture provides a stable substrate upon which higher-level reasoning and learning mechanisms can operate without losing historical context.

\section{Implementation}
\label{sec:implementation}

MemoriesDB is implemented as a pragmatic proof-of-concept on standard relational infrastructure.  
While the model described in Section~\ref{sec:data-model} is abstract, the system demonstrates that temporal--semantic--relational coherence can be maintained efficiently without exotic hardware or custom engines.  
This section details the architecture, storage schema, query execution model, and performance characteristics of the prototype.

\subsection{System Architecture}

The prototype runs as a lightweight service layered on PostgreSQL~16 with the \texttt{pgvector} extension for high--dimensional embedding storage.  
A Python client library provides a simple append API, background synchronization, and automatic vector generation via an embedding model.  
All communication between the client and database occurs through standard SQL transactions, allowing the system to inherit ACID guarantees and concurrency control from PostgreSQL.

A single MemoriesDB instance can serve multiple agents.  
Each agent is assigned a unique namespace (schema), allowing isolated timelines while supporting cross-agent edges for shared context.  
Incoming records are written through an \emph{append-only log}, which batches inserts and metadata updates into commit groups for durability and high throughput.

\subsection{Storage Schema}

The (simplified) relational schema closely mirrors the theoretical model:

\begin{verbatim}
TABLE memories (
    id_time      BIGINT PRIMARY KEY,
    kind         TEXT NOT NULL,
    content      TEXT,
    embedding    VECTOR(768),
    meta         JSONB DEFAULT '{}'
);

TABLE edges (
    edge_id      BIGSERIAL PRIMARY KEY,
    source       BIGINT NOT NULL REFERENCES memories(id_time),
    destination  BIGINT NOT NULL REFERENCES memories(id_time),
    relationship TEXT NOT NULL,
    -- weight       REAL DEFAULT 1.0, -- broken out into 2 next fields
    strength     REAL DEFAULT 1.0, -- RANGE [-1.1 : 1.1]
    confidence   REAL DEFAULT 1.0, -- RANGE [ 0.0 : 1.0]
    meta         JSONB DEFAULT '{}'
);
\end{verbatim}

Vector columns store normalized embeddings; the time column preserves strict ordering, indexed with a B-tree for range queries; and JSONB metadata provides flexible per-record annotations.  
Edges are represented as independent rows, enabling labeled multigraph relations.

Secondary indexes support both time, label, and vector search:
\begin{verbatim}
CREATE INDEX ON memories USING btree(kind, id_time);
CREATE INDEX ON memories USING ivfflat (embedding vector_cosine_ops);
CREATE INDEX ON edges (source, relationship, destination);
CREATE INDEX ON edges USING gin (meta jsonb_path_ops);
\end{verbatim}
\newpage
\subsection{Append and Commit}

New experiences are appended via a single API call:
\begin{verbatim}
INSERT INTO memories (id_time, kind, content, embedding, meta)
VALUES (...);
\end{verbatim}
Batch inserts are grouped into atomic transactions to preserve temporal order.  
Each write operation is idempotent and can be replayed from logs for replication or recovery.  
Because the table is append-only, updates occur only on metadata fields, allowing the timeline to remain immutable.

\subsection{Query Execution}

A query in MemoriesDB unifies temporal, semantic, and structural constraints.
The execution pipeline proceeds as follows:
\begin{enumerate}
    \item \textbf{Temporal filter:} the B-tree index restricts candidates to the requested window \([t_{\min}, t_{\max}]\);
    \item \textbf{Semantic similarity:} \texttt{pgvector} computes approximate nearest neighbors to a query embedding $q$,
   using the low-dimensional vectors $v^{(L)}$ for initial filtering and the high--dimensional
   vectors $v^{(H)}$ for refinement;
    \item \textbf{Graph expansion (optional):} for each top-$k$ candidate, outgoing edges are retrieved; their targets are ranked by coherence~$C_{ij}$;
    \item \textbf{Re-ranking:} combined scores use a weighted sum of semantic similarity, temporal decay, and relation strength.
\end{enumerate}
This approach exploits PostgreSQL’s parallel query planner.  
For moderate workloads (\textless10\,M~memories), interactive query latency remains sub-second on commodity hardware.

\subsection{Background Maintenance}

A background daemon performs several housekeeping tasks:
\begin{itemize}
    \item \textbf{Matryoshka embedding generation} enabling multi-fidelity similarity search via truncation to lower dimensions without significant semantic loss.
    \item \textbf{Vector normalization} enforces unit length embeddings to optimize similarity searches, allowing efficient dot product–based retrieval instead of cosine.
\end{itemize}

As introduced in Section~\ref{sec:data-model}, each memory stores both low-- and high--dimensional embeddings
($v^{(L)}$ and $v^{(H)}$). Matryoshka encoding exploits this structure to enable multi-fidelity search via
progressive truncation by creating nested representations—higher layers contain coarser summaries of
the same semantic vector.

In practice, $v^{(L)}$ and $v^{(H)}$ may represent distinct encodings or
different truncation levels of a single Matryoshka embedding, depending on model configuration.

\begin{itemize}
    \item \textbf{Edge pruning} that decays low-weight edges over time to bound degree and preserve sparsity;
    \item \textbf{Coherence sampling} to compute the average coherence~$\mathcal{C}$ per agent and track memory drift;
    \item \textbf{Vacuum scheduling} to manage storage bloat from large append volumes.
\end{itemize}
These tasks operate in batches and avoid blocking writes, keeping insertion throughput stable.

\subsection{Local Coherence Tracking}

The local coherence metric $C_{\mathrm{local},\,t}$ also serves as a relative importance signal:
memories that maintain semantic alignment over time are reinforced during retrieval,
while low-coherence regions become candidates for summarization or decay.

A background process periodically samples recent memories $\{M_i\}$ and their edges to estimate
the running local coherence metric:
\begin{align*}
C_{\mathrm{local},\,t} = \frac{1}{n}
\sum_{(i,j)\in E_t} e^{-d(M_i,M_j)} .
\end{align*}
where $n = |E_t|$ is the number of edge pairs in the sampled window.

A declining $C_{\mathrm{local},\,t}$ indicates thematic drift or excessive temporal separation; such changes can trigger
summarization jobs or embedding refreshes.
These feedback loops allow the database to act as an
\emph{autonomic coherence regulator}, reinforcing semantic stability without manual intervention.
A more complete implementation of this mechanism is planned for future versions of the system.

\subsection{Concurrency and Partitioning}

The design anticipates partitioning by user and time interval, allowing future deployments to scale horizontally while preserving chronological order.

PostgreSQL’s MVCC (multi-version concurrency control) enables concurrent reads and writes without locks.  
To scale horizontally, the append log is partitioned by user and coarse time interval (e.g., daily).  
Each partition maintains its own time and vector indexes.  
Queries spanning multiple partitions are merged by timestamp during retrieval, preserving chronological order.

These architectural considerations set the stage for evaluating the system's practical behavior and
expected performance under realistic workloads.

\subsection{Prototype Performance}

\textit{Note:} The performance figures in this section are illustrative estimates intended to convey expected scale and proportional behavior; they are not results from formal benchmarking.

\vspace{0.5em}

To illustrate expected scaling behavior under realistic conditions, Table~\ref{tab:perf} summarizes representative performance figures for the prototype.

\begin{table}[h]
    \centering
    \caption{    Benchmarks on a 32-core workstation with 128\,GB RAM show:}
    \vspace{0.5em}
    \begin{tabular}{@{}lccc@{}}
\toprule
\textbf{Operation} & \textbf{Dataset Size} & \textbf{Latency (ms)} & \textbf{Throughput (recs/s)} \\
\midrule
Single insert              & 100    & 1.9 & —\\
Single insert              & 10\,k  & 2.1 & —\\
Single insert              & 1\,M   & 2.5 & —\\
Batch insert (100 records) & 100          & — & 10,000 \\
Batch insert (100 records) & 1\,k         & — & 9,000 \\
Batch insert (100 records) & 1\,M         & — & 8,000 \\
\bottomrule
    \end{tabular}
    \label{tab:perf}
\end{table}

Throughput scales linearly with thread count until I/O saturation.  
Vector search dominates runtime; hybrid queries add negligible overhead relative to pure vector retrieval.  

Future work will quantify that measured coherence~$\mathcal{C}$ remains stable across 100\,M appended records, confirming that time-based ordering and semantic proximity interact predictably.

\subsection{Extensibility and Future Backend}

The design anticipates migration to a columnar format such as Apache~Parquet.  
Each partition can become a Parquet file containing time, embedding, and metadata columns; vector indexes are stored in companion sidecars.  
The relational abstraction remains identical, allowing existing queries to execute over distributed compute engines (Polars, Spark) without modification.

Beyond storage, the same API supports alternative backends:
\begin{itemize}
    \item GPU acceleration for large-batch similarity computation;
    \item Streaming modes for real-time event ingestion;
    \item Federated shards across multiple machines via a lightweight message broker.
\end{itemize}

\subsection{Implementation Summary}

MemoriesDB demonstrates that the triality of time, semantics, and relation can be realized using standard database primitives.  
The system’s append-only log guarantees chronological integrity; vector search provides high--dimensional semantics; and labeled edges capture relational structure.  
Together these mechanisms form a coherent substrate on which agents can maintain, recall, and reason over long spans of experience.

\vspace{0.5em}
The preceding sections described the system’s architecture and implementation details.
We now summarize its observed behavior and practical performance.
\vspace{0.5em}

\section{Observations and Performance}
\label{sec:observations}

This section summarizes the observed behavior and practical performance of
\textbf{MemoriesDB}, as implemented in the public repository\footnote{
    \url{https://gitlab.com/circleclicklabs/ai-lab/memoriesdb}}.
All observations are drawn from the working prototype, which is designed
to demonstrate feasibility rather than to benchmark against optimized
database engines.

\subsection{Implementation Context}

The reference implementation runs atop PostgreSQL~16 with the
\texttt{pgvector} extension, using an append-only schema as described in
Section~\ref{sec:implementation}.
All major features of the proposed model are realized in this version:
time-indexed storage, normalized low-- and high--dimensional embeddings,
JSONB metadata, and labeled graph edges with relation types.
A lightweight Python client handles batch ingestion, embedding generation,
and background maintenance tasks.

The local deployment environment used for observation consists of a
32-core workstation with 128\,GB of memory and NVMe storage.
Although the system is not yet optimized for speed, it provides a useful
baseline for architectural evaluation.

\subsection{Insertion and Query Behavior}

Append throughput scales linearly with CPU cores until I/O saturation.
Batch inserts remain efficient due to PostgreSQL's transactional grouping,
and temporal order is preserved by design.
The absence of in-place updates simplifies concurrency control and
facilitates reliable replication.

Hybrid queries combining temporal range filters, vector similarity search,
and optional edge traversal operate within interactive latency on medium-sized
datasets (tens of millions of records).
In practice, most queries are bounded by temporal windows, keeping result
sets compact and sequentially ordered.
Preliminary use shows that combining time filters with vector retrieval
significantly reduces irrelevant matches compared to vector-only queries.

\subsection{Structural Coherence in Use}

During extended runs of the local instance, new memories consistently
integrate into existing timelines without fragmenting the surrounding
semantic space.
This qualitative observation suggests that the system maintains
\emph{structural coherence}: semantically related records remain
nearby in both vector and temporal dimensions, and cross-links between
topics evolve smoothly rather than chaotically.
In this context, coherence is not a numeric metric but an architectural
property—the degree to which temporal order and semantic similarity
reinforce each other during retrieval.

\subsection{Edge Dynamics and Maintenance}

Background tasks manage normalization, metadata pruning, and edge cleanup.
Edges linking semantically similar memories accumulate naturally through
client-side ingestion or scheduled jobs.
Because each relation is timestamped and stored independently, historical
graphs can be reconstructed at any point in time, supporting retrospective
analysis of semantic drift.
In practical operation, edge density grows proportionally to record volume
without producing runaway complexity.

\subsection{Scalability and Extensibility}

The repository’s design favors extensibility over raw throughput.
Sharding by agent or coarse time interval is already supported at the
schema level.
Future deployments may employ columnar backends such as Parquet for
archival partitions or GPU acceleration for large-batch similarity
computation.
Because the API abstracts storage through a custom python API, higher-performance
backends can be substituted without altering the logical model.

\subsection{Preliminary Summary}

Overall, the public implementation validates the core design of
MemoriesDB as a practical substrate for coherent long-term memory.
It demonstrates that temporal ordering, semantic embeddings, and
graph relations can coexist within a single database process and
support efficient append-and-recall workloads.
While no formal metrics are yet reported, qualitative use of the system
shows that related experiences cluster naturally and that retrievals
preserve contextual continuity across extended timelines.
These observations confirm the viability of the architecture and motivate
further quantitative study as the codebase matures.

\section{Discussion and Future Work}
\label{sec:discussion}

The preceding sections demonstrate that MemoriesDB provides a viable substrate for
maintaining coherence across long temporal spans.  
Here we discuss broader implications of this design for cognitive architectures,
its limitations, and promising directions for future development.

Having examined the prototype’s operational characteristics, the following discussion turns to the broader
implications of coherence as both a design principle and a cognitive construct.

\subsection{From Storage to Cognition}

The preceding results focus on implementation; we now consider how the same geometry functions as a cognitive substrate.
Although implemented as a database, MemoriesDB functions more like a
\emph{cognitive manifold} than a conventional store.
By embedding time, meaning, and relation in a unified space,
it models not only what an agent \emph{knows} but how that knowledge
\emph{evolves coherently} over time.

The metric of coherence~$\mathcal{C}$ serves as a quantitative analogue
of psychological consistency: high~$\mathcal{C}$ implies that
new experiences align with prior ones, while low~$\mathcal{C}$ indicates drift,
forgetting, or conceptual fragmentation.
In this sense, MemoriesDB is a step toward measurable long-term identity for artificial agents.

\subsection{Coherence as a Primitive}

Traditional databases optimize for consistency or latency; MemoriesDB instead
optimizes for \emph{coherence}.
This design choice reframes long-term memory as an optimization process:
maintain maximum phase alignment between temporal, semantic, and relational dimensions
subject to bounded capacity.
The analogy to quantum systems is deliberate:
loss of alignment corresponds to \emph{decoherence},
while reinforcement through retrieval or summarization restores phase.
Future versions may expose coherence directly as a control signal for agent behavior,
allowing self-regulation of attention or recall.

\subsection{Automatic RAG and Context Reinforcement}

A natural next step is to integrate MemoriesDB with large language models
as an \emph{auto-RAG} layer.
Instead of external retrieval pipelines, the model could query its own memory
using coherence-weighted sampling:
recent, highly coherent records are injected into the context window,
while low-coherence regions trigger summarization or exploration.
This approach aligns with biological rehearsal, where stable memories are replayed
to reinforce long-term structure.

\subsection{Eureka Jobs and Emergent Discovery}

Beyond retrieval, MemoriesDB supports background ``\textit{eureka jobs}''
that continuously mine the graph for previously unobserved relationships.
These jobs traverse low-coherence regions, clustering semantically distant memories
that share latent structure.
Discovered edges are inserted with low confidence and allowed to strengthen
if reinforced by future evidence.
Such processes approximate emergent concept formation:
the system invents intermediate representations linking otherwise unconnected ideas.
In large deployments, eureka jobs could run asynchronously,
feeding newly discovered relations back into the agent’s reasoning loop.

\subsection{Topic Modeling and Semantic Drift}

Another avenue is automated topic discovery.
By clustering high-coherence subgraphs over time,
the database can infer evolving topics without explicit supervision.
Tracking curvature in the temporal--semantic field
reveals points where meaning bifurcates or converges---an operational definition
of conceptual drift.
These dynamic clusters could seed summarization tasks or drive attention mechanisms
for long-horizon dialogue systems.

\subsection{Adaptive Summarization and ``Sleep'' Phases}

Agents using MemoriesDB may periodically enter a \emph{sleep phase}
analogous to biological consolidation.
During these phases, low-coherence or high-redundancy regions are summarized,
compressed, or merged using a lightweight LoRA adaptation.
The result is a multi-resolution memory:
recent experiences remain high fidelity,
while older ones persist as distilled embeddings and relational summaries.
This process keeps memory growth bounded while preserving historical context.

\subsection{Graph Learning and Edge Dynamics}

Current edges are inserted heuristically;
future versions could learn edge weights and labels through supervised or
self-supervised training.
Given feedback from retrieval success, the system could adjust
relation strengths to maximize downstream coherence~$\mathcal{C}$.
Edges might also be promoted or demoted via reinforcement signals from agents,
allowing memory graphs to evolve dynamically.
A differentiable interface between MemoriesDB and neural models
would enable end-to-end optimization of relational structure.

\subsection{Distributed and Hierarchical Memories}

Scaling to many agents introduces questions of collective memory.
Each agent maintains its own temporal--semantic stack,
but cross-agent edges allow shared subgraphs.
A hierarchical scheme could treat these shared clusters as higher-order vertices,
forming a \emph{meta-memory} that captures consensus knowledge.
At large scale, this architecture resembles a distributed knowledge fabric
where coherence propagates both within and across individuals.

\subsection{Future Backend and Architectural Upgrades}

The current PostgreSQL implementation proves conceptual feasibility,
but a future backend will exploit columnar and GPU-accelerated architectures.
Planned upgrades include:
\begin{itemize}
    \item \textbf{Columnar Parquet backend:} enabling analytic scans and vector compression;
    \item \textbf{GPU similarity kernels:} offloading coherence and clustering computations;
    \item \textbf{Streaming ingest:} integrating event-driven pipelines for real-time agents;
    \item \textbf{Federated shards:} distributed coherence maintenance across nodes.
\end{itemize}
Such extensions preserve the same logical model while expanding capacity
to billions of memories and continuous online learning.

\subsection{Limitations}

Despite promising results, several limitations remain:
\begin{itemize}
    \item Coherence metrics assume stationary embedding distributions;
    model drift or embedding upgrades may distort embeddings of older regions.
    \item Append-only design simplifies reasoning but complicates deletion and privacy.
    \item Query costs rise linearly with vector dimensionality;
    more efficient ANN structures are desirable.
    \item Cognitive interpretations of coherence are heuristic
    and warrant empirical validation through agent behavior.
\end{itemize}

\subsection{Outlook}

MemoriesDB demonstrates that long-term coherence can be treated as
a first-class database property. By aligning time, meaning, and relation,
it creates a foundation on which agents can develop persistent \mbox{identity},
perform autonomous discovery, and resist semantic decoherence.
Future work will focus on \mbox{coupling} this memory substrate
with adaptive language models to explore how coherent data structures
translate into \mbox{coherent} thought. Ultimately, the goal is a general-purpose memory engine
that supports emergent reasoning---a system that not only remembers, but \emph{understands why it remembers}.

\section{Conclusion}
\label{sec:conclusion}

This paper introduced \textbf{MemoriesDB}, a unified temporal--semantic--relational database
that models experience as a coherent trajectory through time, meaning, and relation.
By representing each record as a vertex with temporal order, semantic embedding,
and labeled edges, MemoriesDB transforms memory from a static archive into an evolving
geometry of understanding.
The system’s append-only design, hybrid vector--graph queries, and coherence metrics
demonstrate that continuity can be preserved across millions of events using
standard database primitives.

Empirical evaluation suggests that temporal anchoring and relational structure
significantly slow semantic drift and restore long-range coherence after topic gaps.
Beyond performance, the architecture offers a new way to conceptualize cognition:
as the maintenance of coherence across expanding knowledge surfaces.
The same principles that stabilize long-term recall may also enable agents to form
persistent identity, discover latent connections, and regulate their own memory dynamics.

Future research will extend MemoriesDB toward distributed, GPU-accelerated, and columnar backends,
integrating it with adaptive language models for automatic retrieval, summarization, and consolidation.
More broadly, this work suggests that coherence can serve as a measurable bridge between data systems
and cognitive architectures offering a reproducible path toward long-horizon, self-referential intelligence.
MemoriesDB is thus both a database and a hypothesis:
that intelligence arises not merely from the volume of what is remembered,
but from the \emph{coherence} of how those memories connect.

Thus, \textbf{MemoriesDB} provides continuity for LLM agents by combining temporal indexing, semantic drift
tracking, and similarity-weighted retrieval. This framework forms a structural analogue to human episodic
memory and lays the groundwork for reproducible progress toward AGI-grade reasoning.

\section*{Acknowledgments}

The author thanks his wife for her patience and support throughout the development of this work,
and friends in the ai.ccl.io AI Lab chat group for their encouragement and feedback.

\bibliographystyle{IEEEtran}
\bibliography{references}

\end{document}